\documentclass[aps,nofootinbib,11pt]{revtex4}

\usepackage{epsf,epsfig,subfigure,graphicx,amsmath,amssymb}
\usepackage{color}
\usepackage{hyperref}

\begin{document}

\title{\bf Bounds on dark matter interpretation of Fermi-LAT GeV excess}

\author{
Kyoungchul Kong$^{(a)}$\footnote{email: kckong@ku.edu} and Jong-Chul Park$^{(a, b)}$\footnote{email: log1079@gmail.com}
}
\affiliation{$^{(a)}$
Department of Physics and Astronomy, University of Kansas, Lawrence, KS 66045, USA
\\
$^{(b)}$ Department of Physics, Sungkyunkwan University, Suwon 440-746, Republic of Korea
}

\begin{abstract}

Annihilation of light dark matter of $m_{\rm DM} \approx (10-40)$ GeV into the Standard Model fermions has been suggested as a possible origin of
the gamma-ray excess at GeV energies in the Fermi-LAT data.
In this paper, we examine possible model-independent signatures of such dark matter models in other experiments such as AMS-02, colliders, and cosmic microwave background (CMB) measurements.
We point out that first generation of fermion final states is disfavored by the existing experimental data.
Currently AMS-02 positron measurements provide stringent bounds on cross sections of dark matter annihilation into leptonic final states,
and $e^+e^-$ final state is in severe tension with this constraint, if not ruled out.
The $e^+e^-$ channel will be complementarily verified in an early stage of ILC and future CMB measurements.
Light quark final states ($q\bar q$) are relatively strongly constrained by the LHC and dark matter direct detection experiments even though these bounds are model-dependent.
Dark matter signals from annihilations into $q\overline{q}$ channels would be constrained by AMS-02 antiproton data which will be released in very near future.
In optimistic case, diffuse radio emission from nearby galaxy (clusters) and the galactic center might provide another hint or limit on dark matter annihilation.

\end{abstract}


\maketitle

\section{Introduction}

The identity of dark matter (DM) is one of the most profound mysteries in particle physics and cosmology.
Various observations of its gravitational effects on multiple scales all point consistently to the existence of dark matter.
However all known particles are excluded as a dark matter candidate and
its identity still remains unknown \cite{DM-Review}.
This situation makes dark matter puzzle as the most pressing motivation for new physics beyond the standard model (SM),
stimulating a variety of searches such as direct, indirect, and collider signatures.
No firm detection has been achieved yet, but several tantalizing hints have been reported.

Numerous DM search experiments have been carried out to observe direct signals by DM scattering off nuclei. DM direct detection experiments such as CDMS-Si~\cite{CDMS-Si}, CoGeNT~\cite{CoGeNT}, CRESST-II~\cite{CRESST-II}, and DAMA/LIBRA~\cite{DAMA} have reported
observations of potential DM events.
However, these signals are not be accepted as significant evidence for DM scatterings due to null results from other experiments including KIMS~\cite{KIMS}, XENON100~\cite{XENON100}, and LUX~\cite{LUX}.

As a complementary method to DM direct detection, indirect detection techniques have been dramatically improved in the last several years, which aim to find signals by DM annihilation and/or decay to SM particles.
Recently, anomalous signals have been reported by many experiments including PAMELA, AMS-02, Fermi-LAT, SPI/INTEGRAL, and XMM-Newton: excesses in the cosmic-ray positron fraction~\cite{PAMELA,AMS}, 130 GeV~\cite{130GeV1,130GeV2} and 511 keV~\cite{511keV} $\gamma$-rays from the galactic center (GC), and X-ray emission around $E_\gamma \simeq 3.5$ keV~\cite{7keV1,7keV2} detected in galaxy clusters.
Recent analyses~\cite{Hooper1, Hooper2, Hooper3, Abazajian:2012pn, Hooper4, Gordon:2013vta, Huang:2013pda, Abazajian:2014fta, HooperNew, Silk} based on the data from Fermi-LAT showed an excess at energies around $1-3$ GeV in the gamma-ray spectrum coming from around the GC, which is consistent with the emission expected from DM annihilations. Among various anomalous indirect signals, this GeV gamma-ray excess is especially interesting since statistical significance of this excess has been gradually increasing with more data from the Fermi-LAT and angular distribution is in good agreement with what is expected from annihilating DM.

Astrophysical uncertainties associated with the extraction of excess in gamma-rays from around the GC have been well discussed including modeling of background emission in the inner galaxy in Ref.~\cite{HooperNew}.
In addition, other possible explanations for the GeV gamma-ray excess have been suggested: a population of millisecond pulsars~\cite{Hooper2, Hooper3, Abazajian:2012pn, Gordon:2013vta, Abazajian:2014fta, Abazajian:2010zy} and pions from the collision of cosmic-rays with gas~\cite{Hooper2, Hooper3, Abazajian:2012pn, Gordon:2013vta}.
In Ref.~\cite{Hooper:2013nhl}, however it was found that the spectral shape from millisecond pulsars is too soft at sub-GeV energies compared to the observed spectrum of the GeV excess and millisecond pulsars can produce no more than $\sim 10\%$ of the gamma-ray excess even including sources known to be millisecond pulsars and unidentified sources which could be pulsars.
Moreover, in Ref.~\cite{HooperNew} it was pointed out that the GeV gamma-ray signal is spatially extended to more than $\sim 10^\circ$ from the GC well beyond the confines of the central stellar cluster which could contain numerous millisecond pulsars.
The analyses of Refs.~\cite{Linden:2012iv, Macias:2013vya} showed that observed distributions of gas provide a poor fit to the morphology of the GeV signal, which moreover cannot account for the spatial extension of the signal~\cite{HooperNew}.

The focus of this study is to investigate implication of the Fermi-LAT gamma-ray signals in other experiments such as AMS-02, PAMELA, Planck, and colliders, assuming DM interpretation is correct.
We present in a single figure a collection of existing bounds by recasting results from various experiments.
Main purpose of our work is to provide a useful overview and guideline in DM model building for the GeV gamma-ray excess including all these bounds which, we think, deserve more attention.
Our recast-process requires appropriate rescaling as well as mapping constraints into a relevant parameter space ($m_{\rm DM}$, $\langle\sigma v\rangle$).
In our analysis, we try to be as model-independent as possible\footnote{In this work, we use a term ``model independent'' in the sense that we provide constraints on the DM annihilation cross sections $\sigma v$ as a function of DM mass $m_{\rm DM}$ for each final state without considering details of annihilation mechanism. However in the case of collider limits, effective operators are used since we need to assume a certain production mechanism for analysis, and thus collider limits have model-dependence.}.
We choose annihilations of dark matter into $\ell\bar\ell$ and $b\bar b$ final states as our reference, and present results in a ($m_{\rm DM}$, $\langle\sigma v\rangle$) plane.
Other scenarios such as democratic annihilations into all kinematically accessible SM fermions and annihilations proportional to $m_f^2$ are also discussed in Section IV.
In most cases such as $\ell\bar\ell$ and $b\bar b$ final states,  recast-process is straightforward and results are easy to convert.
For complicated final states in Section IV, we rescale the limits by considering the corresponding annihilation fraction and characteristics of each final state.
In the case of LEP and LHC bounds, we recompute $\langle\sigma v\rangle$ ourselves with limits on the cutoff scale $\Lambda$ obtained from a literature.
We begin our discussion in Section II by reviewing the Fermi-LAT GeV gamma-ray excess.
We, then, consider various constraints in Section III. Section IV is reserved for discussion.

\section{Fermi-LAT GeV gamma-ray excess}

A gamma-ray excess at GeV energies around the GC has been identified in the Fermi-LAT data by several groups.
\cite{Hooper1, Hooper2, Hooper3, Abazajian:2012pn, Hooper4, Gordon:2013vta, Huang:2013pda, Abazajian:2014fta}.
In Ref.~\cite{HooperNew}, authors reexamined the gamma-ray emission with high resolution gamma-ray maps which was obtained by applying cuts to the Fermi-LAT event parameter CTBCORE and suppressing the tails of the point spread function. In the analysis, they confirmed a significant GeV gamma-ray excess with a spectrum and morphology in close agreement with the expectations from DM annihilation, which was very well fitted by $30-40$ GeV DM particles annihilating to $b\overline{b}$ with an annihilation cross section of $\langle\sigma v\rangle = (1.4-2.0) \times 10^{-26} {\rm cm}^3/{\rm s}$ for a generalized Navarro-Frenk-White (NFW) halo profile with an inner slope of $\gamma=1.26$ and a local DM density of $\rho_\odot = 0.3\, {\rm GeV}/{\rm cm}^3$. With further investigation, it was found that the angular distribution of the excess is approximately spherically symmetric and centered around the dynamical center of the galactic plane. In addition, they observed that the signals are extended to more than $10^\circ$ from the GC, and thus the possibility that millisecond pulsars are responsible for the excess is disfavored.

In Refs. \cite{Gordon:2013vta, HooperNew}, it was also shown that a DM mass of $\sim$ 10 GeV is required when DM annihilates into lepton pairs but the fit to the data favors the case of a DM mass of $30-40$ GeV with a pure $b\overline{b}$ final state. Authors of Ref.~\cite{Silk} pointed out that
a contribution of the diffuse photon emissions originating from primary and secondary electrons produced in DM annihilations is quite significant especially for leptonic final states ($\ell \bar \ell$), which was however neglected in the literature, while such contributions are sub-dominant for the $b\overline{b}$ channel.  Considering the inverse Compton scattering and Bremsstrahlung contributions from electrons, they found that annihilations of $\sim 10$ GeV DM particles into the purely leptonic final state provide a little better fit than the pure $b\overline{b}$ final state.
In addition, it was shown that 10 GeV DM democratically annihilating into pure $\ell\bar\ell$ final states provides the best $\chi^2$ fit for an annihilation cross section of $\langle\sigma v\rangle = 0.86 \times 10^{-26} {\rm cm}^3/{\rm s}$ for a generalized NFW halo profile with $\gamma=1.2$ and $\rho_\odot = 0.36\, {\rm GeV}/{\rm cm}^3$ and 30 GeV DM annihilating into pure $b\overline{b}$ states does for $\langle\sigma v\rangle = 2.03 \times 10^{-26} {\rm cm}^3/{\rm s}$. Note that ``democratic annihilation into pure $\ell \bar \ell$ states'' implies equal annihilation cross sections into each of $e^+e^-$, $\mu^+\mu^-$, and $\tau^+\tau^-$ channels.

\begin{figure}[t]
\begin{center}
\includegraphics[width=0.49\linewidth]{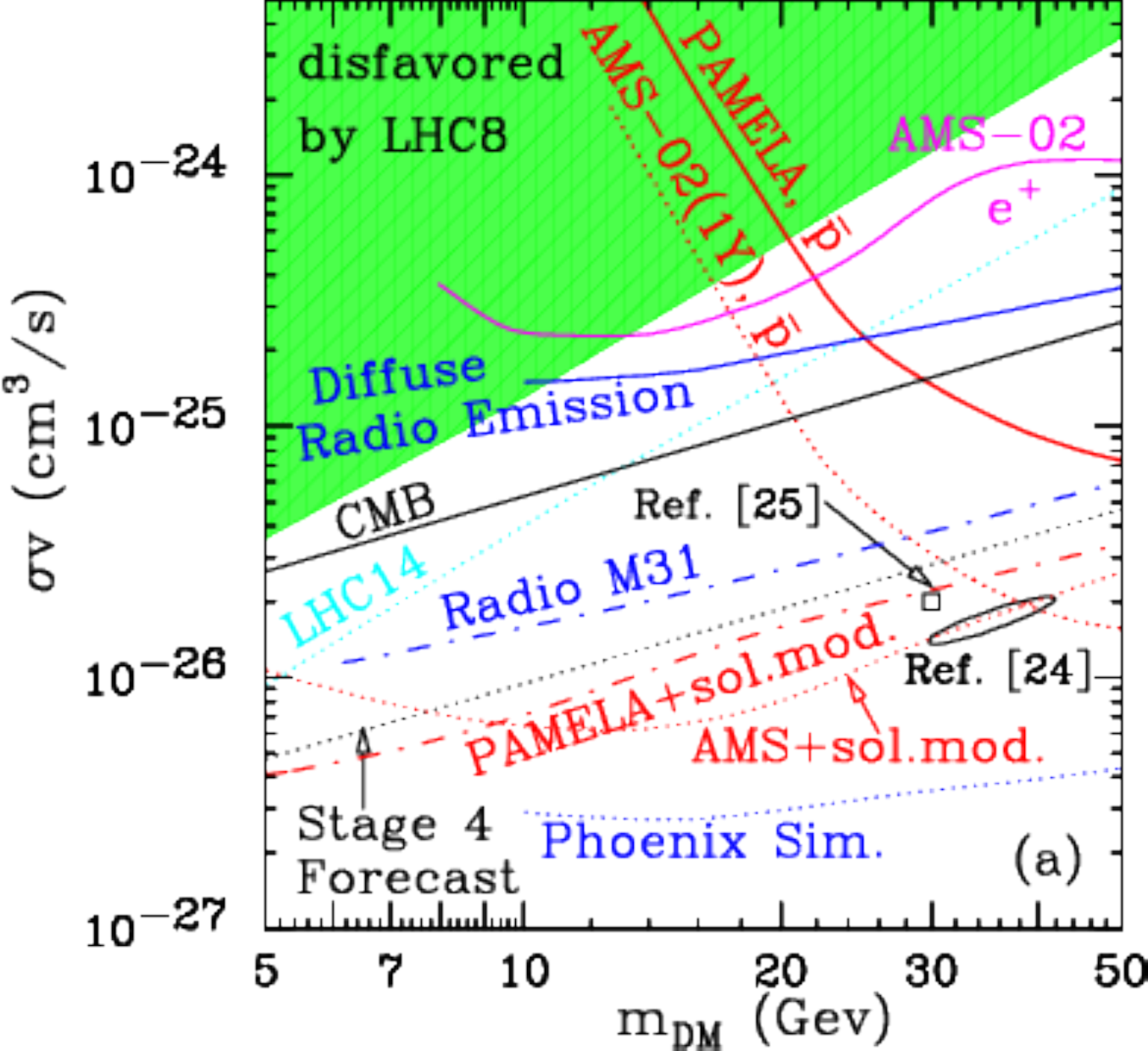}
\includegraphics[width=0.49\linewidth]{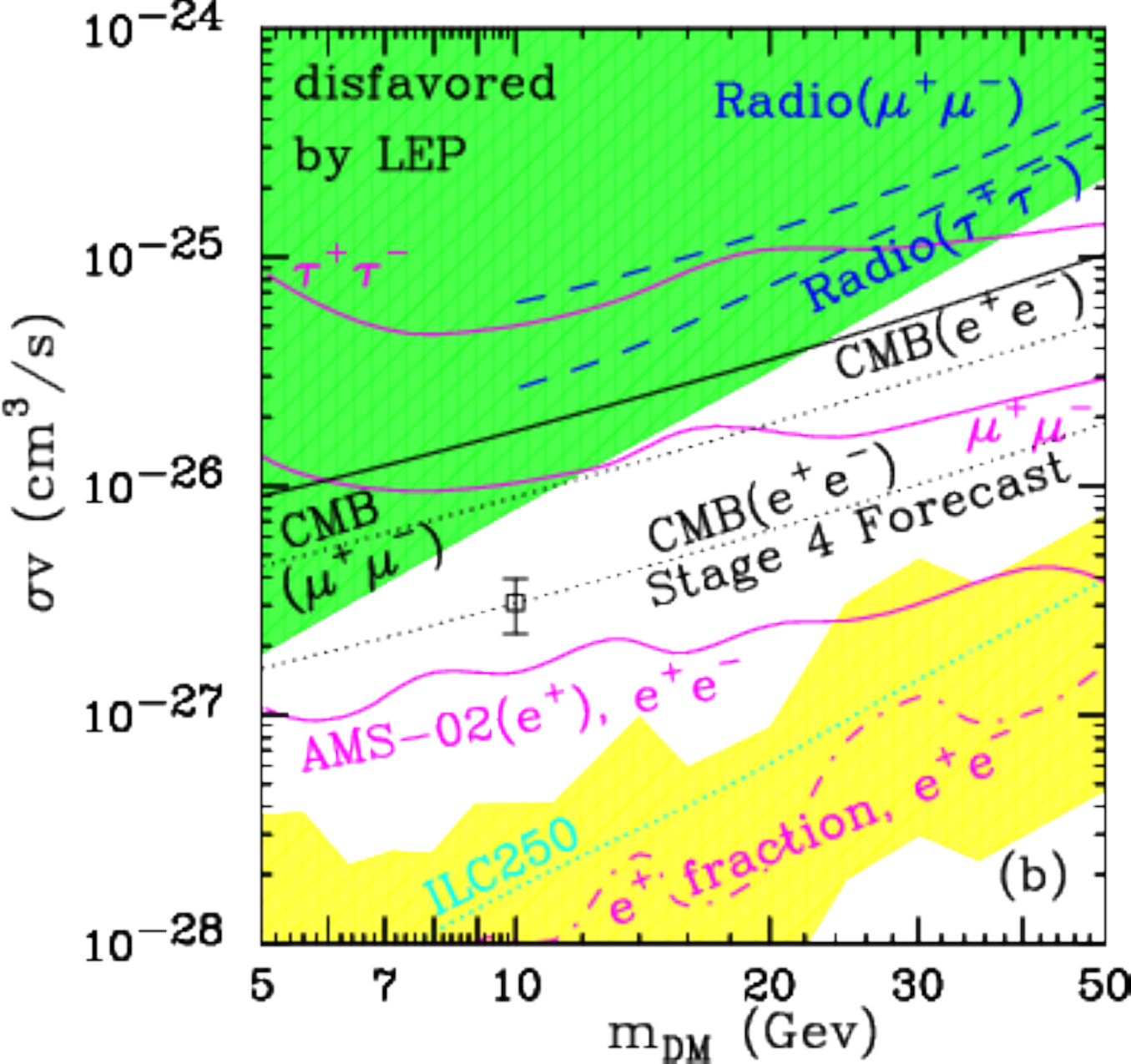}
\end{center}
\vspace*{-0.7cm}
\caption{
Various constraints on dark matter annihilation cross section as a function of dark matter mass
in the final states with (a) $b\bar{b}$ and (b) $\ell\bar\ell$.
We present limits coming from antiproton flux (in red),
diffuse radio emission (in blue), CMB (in black),
positron data (in magenta), and colliders (green-shaded regions).
The yellow shaded band indicates the uncertainties from the local DM density and $e^\pm$ energy loss rate for the AMS-02 positron fraction limit (magenta dot-dashed curve).
Current bounds are shown in solid, dashed or dot-dashed curves while the projected sensitivities are denoted by dotted curves.
Two reference cross sections marked as `square' are fitted results in Ref. \cite{Silk}.
A rescaling factor of $1/3$ is taken into account for democratic annihilations into leptons as shown in (b).
Therefore the same cross section is applied to each leptonic final state.
We also show the 3$\sigma$-contour in (a) from Ref.~\cite{HooperNew}.
}
\label{Fig1}
\end{figure}

We use the best-fit values from Ref.~\cite{Silk} as reference points in our study: $\langle\sigma v\rangle_{\ell \bar\ell} = 0.86 \times 10^{-26} {\rm cm}^3/{\rm s}$ with $m_{\rm DM} = 10$ GeV and $\langle\sigma v\rangle_{b\overline{b}} = 2.03 \times 10^{-26} {\rm cm}^3/{\rm s}$ with $m_{\rm DM} = 30$ GeV. As discussed in Ref. \cite{Silk}, the diffusion model induces an additional uncertainty, which is quantified by the MIN, MED, and MAX sets of propagation parameters (see Ref.~\cite{Donato:2003xg}.). Thus, the uncertainty on the diffusion model parameter sets is converted into an error on the best-fit value for the cross section: $\langle\sigma v\rangle_{\ell \bar \ell} = (0.68-1.18) \times 10^{-26} {\rm cm}^3/{\rm s}$, which is shown as a vertical bar in Figure~\ref{Fig1} (b). We also include the best-fit range of the DM mass and annihilation cross section for the pure $b\overline{b}$ final state obtained in Ref.~\cite{HooperNew} as a (black) contour in Figure~\ref{Fig1} (a). In the next section, we will study possible constrains on the DM annihilation cross sections for each annihilation channel, and for an easier comparison the rescaled best-fit values of $\langle\sigma v\rangle_{\ell \bar\ell}$ by a factor of $1/3$ is therefore presented in Figure~\ref{Fig1} (b) due to the assumption of {\it democratic annihilations into leptons}
as in Ref. \cite{Silk}.

In Refs.~\cite{HooperNew, Silk}, it was shown that
10 GeV DM democratically annihilating into $\ell \overline{\ell}$ final state
and $20-40$ GeV DM into $q\overline{q}$ provide
the best $\chi^2$ fits to the spectrum of the GeV gamma-ray excess, which are therefore used as reference points in our analysis.
In the case of quark final state, we particularly choose $30-40$ GeV DM with $b$-quark final to be more conservative since light quarks are more strongly constrained by LHC and antiproton observations.
In addition, the authors of Ref.~\cite{HooperNew} found that
democratic annihilations into all kinematically accessible SM fermions ($m_{\rm DM}\simeq 18-26$ GeV) and annihilations proportional to $m_f^2$ ($m_{\rm DM}\simeq 28-39$ GeV) also provide a good-fit.
Consequently, we will also discuss those possibilities in Section IV.

\section{Constraints}

In a large class of DM models, the annihilations or decays of DM particles can produce various cosmic-ray fluxes such as $e^{\pm}, \overline{p}, \gamma$, and $\nu$ which are possibly observed or constrained by cosmic-ray experiments.
Moreover, depending on the mass and interactions of DM particles they can be produced in colliders and/or leave signals in DM direct detection experiments. If DM annihilations into $\ell \bar \ell$ and/or $b\overline{b}$ are really responsible for the observed Fermi-LAT GeV gamma-ray excess, such annihilation channels would be therefore constrained by various DM searches.
In order to explain the GeV excess and direct search results including CDMS-Si and CoGeNT in a single framework with $\sim 10$ GeV DM, the authors of~\cite{Kyae:2013qna} have generally explored various DM annihilation and scattering processes discussing important phenomenological constraints coming from particle physics.
Refs.~\cite{Alves:2014yha, Berlin:2014tja, Izaguirre:2014vva} discussed possible constraints on $\sim 35$ GeV DM annihilating into $b\overline{b}$ and $\sim 25$ GeV DM annihilating democratically into SM fermions from the LUX and the LHC and detection prospects in near future direct detection experiments. Along this line, $b$-quark flavored DM models were suggested in Refs.~\cite{Boehm:2014hva, Agrawal:2014una}.

As a complementary study, in this paper we discuss current and future model-independent bounds from positron, antiproton, CMB, radio emission, neutrino measurements, DM direct detection experiments, and collider experiments.

%
%%%%%%%%%%%%%%%%%%%%%%%%%%%%%%%%%%%%%%%%%%%%%%%%%%%%%%%%%%%%
%%%%%%%%%%%%%%%%%%%%%%%%%%%%%%%%%%%%%%%%%%%%%%%%%%%%%%%%%%%%
%

\subsection{Indirect detection}

\subsubsection{Positron}

Exquisite measurements of the cosmic-ray positron flux as well as the positron fraction, recently performed by AMS-02~\cite{AMS}, have allowed us to set bounds on the DM annihilation cross sections or decay lifetime to the SM particles since the annihilation or decay of the DM particles produces a positron flux.
Especially, excellent agreement of low energy positron measurements with the expected standard backgrounds provides very stringent upper limits on the annihilation cross sections $\langle\sigma v\rangle$ for various DM annihilation channels including  $e^+e^-$, $\mu^+\mu^-$, $\tau^+\tau^-$, and $b\overline{b}$ \cite{Bergstrom,Ibarra}.
We do not intend to explain the AMS-02 data in terms of DM annihilation but
to use it to constrain the relevant parameter space ($m_{\rm DM}$, $\langle\sigma v\rangle$).

In Ref.~\cite{Ibarra}, for the positron background authors used the widely-accepted assumption that the background contains mainly two components: (i) a secondary positron produced in collisions of primary cosmic rays in the interstellar medium with a simple power-law spectrum and (ii) possibly a primary positron component from astrophysical sources such as pulsars approximated by a power-law with an exponential cut-off. To include solar modulation effects at low energies, the authors additionally estimated the flux at the top of the atmosphere relating to the flux at the heliospheric boundary under the force field approximation, and then performed a $\chi^2$ test with this pure background model for the AMS-02 positron data in the range of $E=2-350$ GeV. They introduced a contribution to the positron flux from the DM annihilations with $m_{\rm DM}$ and $\langle\sigma v\rangle$ and recalculated the $\chi^2$ of the best fit model. Then the $2\sigma$ exclusion line is obtained by determining, for a given final state and $m_{\rm DM}$, $\langle\sigma v\rangle$ providing a $\chi^2$ which exceeds that of the pure background model by more than 4.
To obtain the positron flux around the earth, they considered the propagation of positrons in the Milky Way, which is usually described by a stationary two-zone diffusion model with cylindrical boundary conditions.
In the analysis, they used the Einasto profile with $\rho_\odot = 0.39\, {\rm GeV}/{\rm cm}^3$ for the density of DM particles in our galaxy halo and the MED model set for propagation parameters of the model sets proposed in Ref.~\cite{Donato:2003xg}.
Following the same procedure, they also obtained limits on $\langle\sigma v\rangle$ using the AMS-02 positron fraction data and showed that the limits from the positron fraction are comparable or stronger by a factor of a few, especially in low DM mass region, than those from the positron flux.
They also derived the limits for the NFW profile and the isothermal profile as well as for the MIN and MAX propagation parameter sets, and showed that the derived limits are mildly (less than $20-30\, \%$) affected by the choice of parameters.

In Ref.~\cite{Bergstrom}, the authors derived limits from the observation that a DM signal would leave a sharp spectral feature in the AMS-02 positron fraction data using the same phenomenological astrophysical background model as in Ref.~\cite{Ibarra}.
In this analysis, they used the Einasto profile with $\rho_\odot = 0.4\, {\rm GeV}/{\rm cm}^3$.
To reflect uncertainties by the $e^\pm$ energy loss rate and local DM density, the authors considered a range of local radiation and magnetic field energy densities $U_{\rm rad} + U_B=(1.2-2.6)\, {\rm eV}/{\rm cm}^3$ and a local DM density $\rho_\odot = (0.25-0.7)\, {\rm GeV}/{\rm cm}^3$,
which is shown as a yellow shaded band in Figure~\ref{Fig1} (b).
For the AMS-02 positron fraction data, the limits from Ref.~\cite{Bergstrom} are stronger by a factor of $2-3$ in low DM mass region ($m_{\rm DM}\lesssim 50$ GeV) than those from Ref.~\cite{Ibarra}, which is mainly due to the difference between the used data ranges: $E=1-350$ GeV versus $E=2-350$ GeV.

In our analysis, we use the $2\sigma$ exclusion limit obtained from the AMS-02 positron flux data for the Einasto profile with the MED propagation model in Ref.~\cite{Ibarra} as a conservative bound, which appears as a magenta solid line for each final state in Figure~\ref{Fig1}.
We also include the upper limit ($95\%$ CL) on the $e^+e^-$ final state from the AMS-02 positron fraction data in Ref.~\cite{Bergstrom} as a magenta dot-dashed line which can vary by a factor of $4-5$ as indicated by yellow shaded band depending on the local DM density and energy loss rate.
The limit on DM annihilation into the $\mu^+\mu^-$ $(\tau^+\tau^-)$, as derived from the AMS-02 positron fraction data in Ref.~\cite{Bergstrom}, is more stringent by a factor of roughly $4-13$ ($4-11$) than that from the AMS-02 positron flux data in Ref.~\cite{Ibarra}.
As can be seen from the right panel of the figure, the $e^+e^-$ annihilation channel is already strongly constrained by the current AMS-02 positron measurements.

\subsubsection{Antiproton}

Antiproton production from DM annihilations or decays is generic in DM models for hadronic or gauge boson channels. Leptonic channels are relevant for $m_{\rm DM} \gtrsim \mathcal{O}(100)$ GeV since antiprotons are mainly produced through electroweak corrections for these channels. Thus, the current precise measurements on $\overline{p}$ by PAMELA~\cite{Adriani:2010rc} and the upcoming results from AMS-02 can constrain such DM annihilation channels.
Ref.~\cite{AntiProton} provided current limits on DM annihilation cross sections for various annihilation channels including $e^+e^-$, $\mu^+\mu^-$, $\tau^+\tau^-$, $b\overline{b}$, and gauge bosons based on the PAMELA antiproton data, as well as the sensitivity of early AMS-02 antiproton measurements.
Ref.~\cite{AntiProton} used data whose kinetic energy is
above 10 GeV to minimize the effect of solar modulation.
Therefore the limits are weaker in the low mass region and even show a rise below $m_{\rm DM} \approx$ 50 GeV.
However, Ref.~\cite{Fornengo:2013xda} includes the solar modulation effect, modeling antiproton propagation in the heliosphere,
and consequently the limits are extended to the low mass region without showing the rise.
Similar results are obtained from BESS-Polar II data taking into account the solar modulation effect~\cite{Kappl:2011jw}.
In our analysis,
we include the results from Ref.~\cite{AntiProton} as conservative bounds as well as more stringent limits from Ref.~\cite{Fornengo:2013xda}.
For a fixed DM mass $m_{\rm DM}$ and annihilation cross section $\langle\sigma v\rangle$, the astrophysical background by standard cosmic-ray processes is optimized within the uncertainty bandwidth in order to minimize the $\chi^2$ of the total $\overline{p}$ flux including the DM annihilation contributions. Authors of Ref.~\cite{AntiProton} assumed the Einasto profile with $\rho_\odot \simeq 0.3\, {\rm GeV}/{\rm cm}^3$ and the MED model of the propagation parameter sets proposed in Ref.~\cite{Donato:2003xg} as a representative model to find the limits on $\langle\sigma v\rangle$. They also showed that the limits are almost same for the NFW profile, $2-3$ times weaker for the Burkert profile, $\sim 10$ times weaker for the MIN propagation model, and $2-3$ times more stringent for the MAX model.
Authors of Ref.~\cite{Fornengo:2013xda} mainly used the Einasto profile with $\rho_\odot \simeq 0.39\, {\rm GeV}/{\rm cm}^3$ and the MED model of the propagation parameter sets.
It was also shown that the limits are slightly weaker for the NFW profile, $\sim2$ times weaker for the cored isothermal profile, $10-15$ times weaker for the MIN model, and $5-10$ times more stringent for the MAX model.
Moreover, they showed that the limits depend weakly on the solar modulation modeling compared to the propagation modeling.

In our study, we take the $2\sigma$ PAMELA exclusion bounds calculated for the Einasto profile with the MED propagation model in Refs.~\cite{AntiProton, Fornengo:2013xda}, which are respectively presented as red solid and dot-dashed curves in Figure~\ref{Fig1} (a).
As shown in Figure~\ref{Fig1} (a), the 1 year of AMS-02 data (red dotted curve), soon-to-be-released, will improve the limits by a factor of $\sim 5$~\cite{AntiProton} ($\sim 2$~\cite{Fornengo:2013xda}) compared to the PAMELA results, and constrain or probe the preferred parameter space in the $b\overline{b}$ channel.
Note that a very recent antiproton analysis in Ref.~\cite{Bringmann:2014lpa} showed improved limits by a factor of $2-5$ compared to previous limits~\cite{Fornengo:2013xda, Kappl:2011jw}. There are two main reasons for the improvement: (i) they used recent update of PAMELA data~\cite{Adriani:2012paa} and (ii) they employed an improved statistical treatment of the background uncertainties (see Ref.~\cite{Bringmann:2014lpa} for more details).

\subsubsection{CMB}

DM annihilations into SM particles could alter the thermal history of our universe through the injection of energy into the photon-baryon plasma, gas, and background radiation.
The injected energy affects the recombination process and the reionization mechanism of the universe, increasing ionization and atomic excitation of the gas and broadening the last scattering surface.
These changes lead to modifications in the CMB temperature and polarization power spectra, the positions of the TE and EE peaks, and the power of polarization fluctuations at large scales.
In Ref.~\cite{CMB}, updated constraints on the DM annihilation cross section and mass were given combining CMB power spectrum datasets from Planck, WMAP9, ACT, and SPT as well as low redshift measurements from BAO, HST, and supernovae.
For current limits on DM annihilations, authors used the temperature data and four-point lensing measurements from Planck.
They also showed that the full Planck temperature and polarization data will improve the current bounds by a factor of $2-3$, and another factor $2-3$ improvement is expected from the proposed CMB Stage IV experiment~\cite{FutureCMB}.
In addition, the authors provided updated deposition efficiency factors $f_{\rm eff, sys}$ considering an updated treatment of the excitation, heating, and ionization energy fractions. The efficiency factor $f_{\rm eff}$ describes the fraction of the injected energy by annihilations of DM particles which is deposited in the plasma.

We choose the $2\sigma$ exclusion limits for $f_{\rm eff}=1$ from Ref.~\cite{CMB} and rescale by the updated $f_{\rm eff, sys}$,
corresponding to each annihilation channel as constraints from CMB observations for our analysis.
In Figure~\ref{Fig1}, the current and projected (CMB Stage IV experiment) constraints from CMB measurements are shown as solid and dotted black lines, respectively. The current limit on the $\mu^+\mu^-$ channel can be easily obtained rescaling the limit on the $e^+e^-$ channel by a factor of $\sim 3$.
In the figure, the limits on the $\tau^+\tau^-$ channel are omitted
since they are almost the same as those on the $\mu^+\mu^-$ channel.
As already stated earlier, the full Planck data release will provide $2-3$ times more stringent limits than the current ones.

\subsubsection{Radio emission}

Relativistic electrons and positrons produced by DM annihilation lose their energy via synchrotron radiation in the presence of magnetic fields.
Such a signal may be produced in nearby galaxy clusters, which are the most massive virialized objects in the universe.
About $80\%$ of the mass of clusters is comprised of DM, making them good candidates for astrophysical searches for a signature from DM.
In Ref.~\cite{Radio}, authors calculated bounds on DM annihilation cross sections using upper limits on the diffuse radio emission or low levels of observed diffuse emission from selected nearby galaxy clusters, or detections of radio mini-halos.
They presented upper limits on $\langle\sigma v\rangle$ for four different annihilation channels, $\mu^+\mu^-\,, \tau^+\tau^-\,, b\overline{b}\,,$ and $W^+W^-$, with a smooth NFW DM profile for two representative galaxy clusters of A2199 and Ophiuchus. The results for two different clusters are comparable to each other.
Effects of uncertainty in the cluster mass and magnetic field parameters were also studied for A2199, which showed uncertainties in the annihilation cross section of a factor of $\sim 2$.
Radio constraints on DM annihilation in the galactic halo~\cite{RadioGC1, RadioGC2, RadioGC3} are generally similar to those from nearby galaxy clusters.
However, those limits critically depend on magnetic field and cosmic-ray diffusion around the galactic center.
Radio emissions from dwarf spheroidal galaxies provide additional constraints which however suffer from unknown magnetic field in dwarf galaxies~\cite{Spheroidal1, Spheroidal2}.

It is known that galaxy clusters may host various subhalos in the mass range of $(10^{-6} - 10^{7} ) M_\odot$\footnote{Depending on DM models, i.e. the DM mass and its coupling to the cosmic background particles, the range of subhalo masses can be extended to  $(10^{-11} - 10^{10} ) M_\odot$~\cite{Profumo:2006bv, Bringmann:2009vf, Bringmann:2011ut}.}, where $M_\odot$ is the solar mass,
and the radio emission limits on DM annihilations strongly depend on the assumed amount of cluster substructure~\cite{Radio}.
The amount of substructure is of great importance since the radio emission flux due to DM annihilation is proportional to the $J$ factor which is in turn proportional to the DM density squared.
The $J$ factor is defined as the line-of-sight integral of the DM density squared:
\begin{eqnarray}
J = \int_{\Delta\Omega} d\Omega \int_{l.o.s.} \rho_{\rm DM}^2(l)\, dl\,,
\end{eqnarray}
where $\Delta\Omega$ is the angular size of the emission region,
can be enhanced by one or more orders of magnitude compared to a smooth NFW profile.
In Ref.~\cite{Radio}, it was shown that a substructure model based on the results of the Phoenix Project\footnote{The Phoenix Project is a series of DM  simulations of different galaxy clusters following the evolution of cluster-sized halos~\cite{Phoenix}.} yields almost two orders of magnitude stringent limits than those by a smooth NFW profile since the $J$ factor for the Phoenix simulation, $J_{\rm Phoenix} = J_{\rm NFW} + J_{\rm sub}$ is dominated by the substructure contribution $J_{\rm sub}$.

For our presentation, we choose the limits for the A2199 cluster with a smooth NFW profile, as shown as blue curves in Figure~\ref{Fig1}
(solid for $b\bar b$ and dashed for $\tau^+ \tau^-$ and $\mu^+\mu^-$).
This result is enhanced by a factor of 2 ($\mathcal{O}(10\%)$), when a substructure model with the cutoff mass $M_{cut} = 10^{-6} (10^7) M_\odot$ is considered.
Limits from the Phoenix Project are shown as the blue dotted curve and are well below the fitted results in the Fermi-LAT gamma-ray signals.
Consequently, diffuse radio emission searches from nearby galaxy clusters might constrain or prove the parameter regions preferred by the Fermi-LAT GeV excess.
A more recent study on radio signals from the neighbor galaxy M31 (Andromeda galaxy)~\cite{RadioM31} indicates that the bound on DM annihilation into the $b\bar{b}$ ($\tau^+\tau^-$) channel can be stronger by a factor of $\sim 6\, (3)$ than that from the A2199 cluster even in the most conservative case.
This limit for the $b\bar b$ final state is shown as a blue dot-dashed curve in Figure~\ref{Fig1} (a).
New analysis on GC radio observations in Ref.~\cite{Bringmann:2014lpa} also provides competitive limits.
However, these limits strongly depend on the core size and inner slope of the DM profile.

\subsubsection{Neutrino}

DM annihilations in the Galactic halo might produce high energy neutrinos which are constrained by IceCube neutrino measurements.
However, these constrains are at the level of $\langle\sigma v\rangle \approx 10^{-22}\, {\rm cm}^3/{\rm s}$ and only applicable for $m_{\rm DM} > \mathcal{O}(100)$GeV due to the low energy limit of IceCube \cite{IceCube}.
The capture and subsequent annihilations of DM particles in the sun would induce neutrino fluxes, which in turn may be observed by neutrino telescopes such as Super-Kamiokande~\cite{SuperK} and IceCube~\cite{IceCubeSun}.
However, these limits are highly model-dependent since the neutrino fluxes from the sun depend on the DM annihilation cross sections as well as DM-nucleus scattering and DM-self scattering cross sections.
Thus, we make no farther discussion on the limits from cosmic neutrino measurements.

\subsection{Direct detection}

DM models fitting the Fermi-LAT GeV excess might be constrained by stringent DM direct detection limits from XENON100~\cite{XENON100} and LUX~\cite{LUX}.
As discussed in Ref.~\cite{Berlin:2014tja}, constraints from DM direct detection experiments are model-dependent, and they are comparable or less stringent compared to the limits from colliders in the missing energy plus $j$/$b$, for most operators.
Therefore we will not consider bounds from direct detection in our analysis.
However certain effective operators such as $\overline{\chi}\gamma^\mu\chi \overline{f}\gamma_\mu f$
are severely constrained even in the absence of couplings to light quarks, {\it i.e.,} with DM couplings to $b$-quarks only \cite{Izaguirre:2014vva}.

\subsection{Collider}

Dark matter pair production at colliders may leave observable signatures in the energy and momentum spectra of the objects recoiling against the dark matter.
Collider limits are complementary to and competitive with limits on dark matter annihilation and on dark matter-nucleon scattering from indirect and direct searches  \cite{Arrenberg:2013rzp}.
These limits, however, do not suffer from systematic and astrophysical uncertainties associated with direct and indirect limits.
We use LEP data on mono-photon events with large missing energy to constrain the coupling of dark matter to electrons,
while we use LHC data on mono-jet search for the coupling of dark matter to $b$ quarks.
Unfortunately for this purpose, we need to assume a certain production mechanism, for which we introduce effective operators.
As can be seen in the following sections, the cut-off scale
$\Lambda$ is $\sim 500$ GeV (from LEP) or larger, which is much larger than the range of DM mass $m_{\rm DM}\simeq10-40$ GeV discussed in this work and final state SM fermion mass $m_f \leq m_b$.
The cut-off scale $\Lambda$ can be considered as a mediator mass scale $M_{\rm med}$ up to couplings to DM and SM fermions.
For couplings of ${\cal O}(1)$, the mediator mass is much larger than the DM mass, and effective operator approach for collider limits is therefore valid.
Note that results from effective operator analysis deviate from exact ones as couplings become smaller.
See Ref. \cite{Fox:2011fx} for more discussion on effective operator approach with light mediators.

\subsubsection{LEP and ILC}

In Ref. \cite{Fox:2011fx}, the following four operators are considered for LEP bounds.
\begin{align}
  \mathcal{O}_V &= \frac{(\bar\chi\gamma_\mu\chi)(\bar \ell \gamma^\mu \ell)}{\Lambda^2} \,,
    & \text{(vector, $s$-channel)} \label{O1} \\
  \mathcal{O}_S &= \frac{(\bar\chi\chi)(\bar \ell \ell)}{\Lambda^2} \,,
    & \text{(scalar, $s$-channel)} \\
  \mathcal{O}_A &= \frac{(\bar\chi\gamma_\mu\gamma_5\chi)(\bar \ell \gamma^\mu\gamma_5 \ell)}{\Lambda^2} \,,
    & \text{(axial vector, $s$-channel)} \label{O2} \\
  \mathcal{O}_t &= \frac{(\bar\chi \ell)(\bar \ell \chi)}{\Lambda^2} \,,
    & \text{(scalar, $t$-channel)} \label{O3}
\end{align}
where $\ell$ and $\chi / \bar\chi$ represent a lepton and a dark matter candidate, respectively.
The $\Lambda$ may be considered as a mass scale of mediator up to couplings to leptons and dark matter.
Two operators, $\mathcal{O}_S$ and $\mathcal{O}_A$, suffer from $s$-wave suppression and it would be difficult to accommodate the Fermi-LAT gamma-ray data,
avoiding collider constraints at the same time.
Therefore in our discussion, only $\mathcal{O}_V$ and $\mathcal{O}_t$ are relevant, and
we choose the vector interaction for illustration since it gives more conservative limit.
For the wide range of dark matter mass, $0 < m_{\rm DM} < 50$ GeV, LEP bounds on the mass scale is found to be $\Lambda \gtrsim 480$ GeV.
With this limit, we use {\tt micrOMEGAs} \cite{Belanger:2013oya} to compute annihilation cross section of $\chi \bar\chi$ into the $e^+ e^-$ final state, $\langle  \sigma v (\chi \chi \to e^+ e^-) \rangle$.
We show LEP constraints as a shaded region (in green) in Figure \ref{Fig1}(b).
A lower bound on $\Lambda$ is equivalent to an upper bound on the annihilation cross section.
Sensitivity of LEP on $\mathcal{O}_t$ operators is better by a factor of 2.
Other operators such as $ \mathcal{O}_{PS} = \frac{(\bar\chi\gamma_5\chi)(\bar \ell \gamma_5 \ell)}{\Lambda^2}$
for \text{(pseudo-scalar, $s$-channel)} are not $s$-wave suppressed and may be considered for the gamma-ray signals.
In this case, the corresponding limits from LEP/ILC are weaker than those with $\mathcal{O}_V$ by a factor of 4 or so \cite{Dreiner:2012xm}.
%%%
Note that Ref. \cite{Fox:2011fx} also provides limits on annihilation cross section, which differ from our results by a factor of 3.
Accounting for three generations, our estimation is in agreement.

As shown in Figure \ref{Fig1}(b), LEP constrains parameter space significantly but still allows (10 GeV, $3\times 10^{-27} {\rm cm^3/s}$).
On the other hand, a future ILC can easily reach  $ \sigma v \sim 3\times 10^{-27} {\rm cm^3/s}$.
Sensitivity of a future ILC has been investigated in Ref. \cite{Chae:2012bq} for $\sqrt s=$ 250 GeV, 500 GeV, and 1 TeV with or without polarization.
In terms of the scale $\Lambda$, the accessible regions are approximately independent of the dark matter particle mass,
until the kinematic reach of the collider (roughly $m_{\rm DM} \approx \sqrt{s}/2$) is reached.
A 250 GeV ILC (the ``Higgs factory'') is sensitive to scales $\Lambda$ up to about 1-1.2 TeV,
a factor of 2.5-3 higher than the LEP bounds given in Ref. \cite{Fox:2011fx}.
The LEP bounds on the annihilation cross section would be improved by one to two orders of magnitude as shown in Figure \ref{Fig1}(b).
A dotted (in light blue) curve labeled by `ILC250' represents the reach of the ILC dark matter searches at the 250 GeV center-of-mass energy
with a luminosity of 250 fb$^{-1}$. The regions above the curves are accessible at the 3-sigma level.
This analysis in Ref. \cite{Chae:2012bq} ignores instrumental backgrounds, and
assumes a systematic error on the background prediction of 0.3\%.
The significance is obtained by combining the statistical and systematic errors in quadratures.

The marked point for democratic annihilations into leptons will be easily accessible within the first year of 250 GeV ILC running.
It is clear that the ILC will be able to probe significantly smaller cross sections: $\langle\sigma v\rangle_{e\overline{e}} \sim 2 \times 10^{-28} {\rm cm}^3/{\rm s}$ for $m_{\rm DM}=10$ GeV in the case of $\mathcal{O}_V$ operators.
Therefore if more than $\sim 3\%$ of the Fermi-LAT gamma-ray signal arises  due to dark matter annihilation into $e^+e^-$ for leptonically annihilating 10 GeV DM, the corresponding signal should be confirmed or excluded at a future ILC.
The sensitivity of ILC250 on $\mathcal{O}_V$ operators is a factor of $\sim 2$ weaker than the limit from the AMS-02 positron fraction data in Ref.~\cite{Bergstrom} as can be seen from Figure \ref{Fig1}(b).
However the sensitivity on $\mathcal{O}_t$ operators is enhanced by a factor of 2, which is thus comparable to the AMS-02 positron fraction bound.
As shown in Figure \ref{Fig1}(b), there is some parameter region that is not firmly excluded by the AMS-02 positron measurements but still within the reach of ILC250 due to the uncertainties by the local DM density and energy loss rate (see yellow shaded band).
In addition, note that the ILC250 curve corresponds to the sensitivity at the $3\sigma$ level but the AMS-02 positron fraction to the upper limit at the $95\%$ CL.

\subsubsection{LHC}

The LHC Collaborations have reported limits on the cross section of $ p p \to \chi \bar \chi + X$,
where $X$ can be a jet, photon, $W$ or $Z$.
In each case, limits are reported in terms of the mass scale $\Lambda$ of the unknown interaction expressed in an effective field theory,
though the limits from the mono-jet mode are known to be the most powerful.
Ref. \cite{Zhou:2013raa} presents extrapolations of the current mono-jet searches at the LHC to potential future hadron collider facilities.
However, when dark matter only couples to heavy flavors such as bottom and top quarks,
the mono-jet search loses sensitivity and the mono-$b$ search becomes more effective \cite{Lin:2013sca}.
We use their results \cite{Zhou:2013raa,Lin:2013sca} and CMS data \cite{CMS:rwa} to constrain the $b\bar{b}$ final state.
As in the previous section, we consider a vector interaction $ \mathcal{O}_V $.

Ref. \cite{Lin:2013sca} uses a scalar-operator for their mono-$b$ study and therefore we compute the ratio of production cross sections
for two different operators, which we found to be approximately 2.
CalcHEP \cite{Belyaev:2012qa} is used for cross section estimation with CTEQ6L and QCD scale = $\sqrt{\hat s}$.
Then we find the limit on the cut-off scale for the vector interaction $ \mathcal{O}_V $ to be
$\sim 650$ GeV after appropriate rescaling with CMS data.
For a dark matter mass between 1 GeV and 50 GeV, current CMS limit on $\Lambda$ is almost constant,
900 GeV at 8 TeV with 19.5 fb$^{-1}$ and 750 GeV at 7 TeV with 5 fb$^{-1}$ from mono-jet searches.
Finally we calculate  $\sigma v(\chi \bar{\chi} \to b \bar{b} )$.
Current CMS bounds are shown as shaded region (yellow) in Figure \ref{Fig1}(a).
Projected limits at 95\% CL on the annihilation cross section at the 14 TeV with a luminosity of 100 fb$^{-1}$ is shown
as a dotted line labeled by `LHC14'.
Our results are consistent with those in Ref. \cite{Zhou:2013raa},
accounting for different luminosity, the number of quark flavors, the reduced cut-off scale from the mono-$b$ search.
Higher luminosity option (3 ab$^{-1}$) in this case only improves sensitivity by a factor of 2-3.

The observed gamma-ray spectrum is generally best fitted by dark matter particles with a mass of $\sim$20-40 GeV,
that annihilate to quarks with a cross section of $\sigma v \sim 10^{-26} cm^3/s$ \cite{HooperNew}.
Therefore LHC14 would disfavor annihilation into the light quark final state, since the final state with $b\bar b$ is already close to the current collider limits.
Due to the nature of collider experiments ($e^+e^-$ and $pp$), their bounds imply that
dark matter annihilation into the heavier lepton flavors would be preferred.
The same is true qualitatively for other effective operators.

\section{Discussion}

Numerous studies identified the GeV gamma-ray excess around the galactic center in the Fermi-LAT data.
This gamma-ray excess could be interpreted as a probable evidence of dark matter annihilations into $b\overline{b}$ with $\langle \sigma v \rangle \approx (1.4-2.1) \times 10^{-26} {\rm cm}^3/{\rm s}$ for $m_{\rm DM} \approx 30-40$ GeV or into $l\overline{l}$ with $\langle \sigma v \rangle \approx (0.6-1.2) \times 10^{-26} {\rm cm}^3/{\rm s}$ for $m_{\rm DM} \approx 10$ GeV.

In this work, we discussed possible model-independent constraints on DM models for the GeV excess.
For leptonic annihilation channels, AMS-02 positron measurements currently provide the most stringent limit on the DM annihilation cross section.
Especially $e^+e^-$ $(\mu^+\mu^-)$ channel is strongly (mildly) constrained by this limit and only small annihilation fraction of $e^+e^-$ channel is allowed suggesting hierarchical annihilation fraction in $e^+e^-$, $\mu^+\mu^-$, and $\tau^+\tau^-$.
Moreover, required parameter space by the $e^+e^-$ channel will be complementarily probed or constrained in an early stage of ILC and future CMB measurements.
In the case of hadronic annihilation modes, light quark channels are relatively strongly constrained by the LHC and DM direct detection experiments, which prefers heavy quark flavors, although these constrains are model-dependent.
For all the quark flavor channels, particulary light quarks, inclusion of the solar modulation effect enhances tension between the GeV gamma-ray excess and the null result in the PAMELA antiproton data.
The AMS-02 antiproton measurement data will further constrain the parameter space or indicate another indirect evidence for DM annihilation.
In an optimistic case with the results of the Phoenix Project, diffuse radio emission searches from nearby galaxy clusters might be able to provide another hint or constraint on the DM annihilation.
It is important to check self-consistency across diffusion and propagation schemes adopted to obtain various constraints such as antiproton and positron limits.
Ref.~\cite{Ibarra} (positron) and Refs.~\cite{AntiProton, Fornengo:2013xda} (antiproton) employed the conventional diffusion and propagation parameter sets (MIN, MED, MAX) from Ref.~\cite{Donato:2003xg} and chose the MED model as a reference.
Ref.~\cite{Bringmann:2014lpa} (antiproton) used the MIN and MAX sets to show uncertainty in different propagation scenarios but chose the KRA model from Ref.~\cite{Evoli:2011id} as their reference.
In Ref.~\cite{Bergstrom} (positron), detailed information on model sets are not given.
For calculation of synchrotron radiation by electrons and positrons, the propagation of cosmic-ray electrons and positrons should be modeled.
Refs.~\cite{RadioGC2, RadioGC3} also used the same (MIN, MED, MAX) parameter sets  as benchmark propagation models for radio constraints on DM annihilations in the galactic halo.

\begin{figure}[t]
\begin{center}
\includegraphics[ width=10cm, height=9.cm]{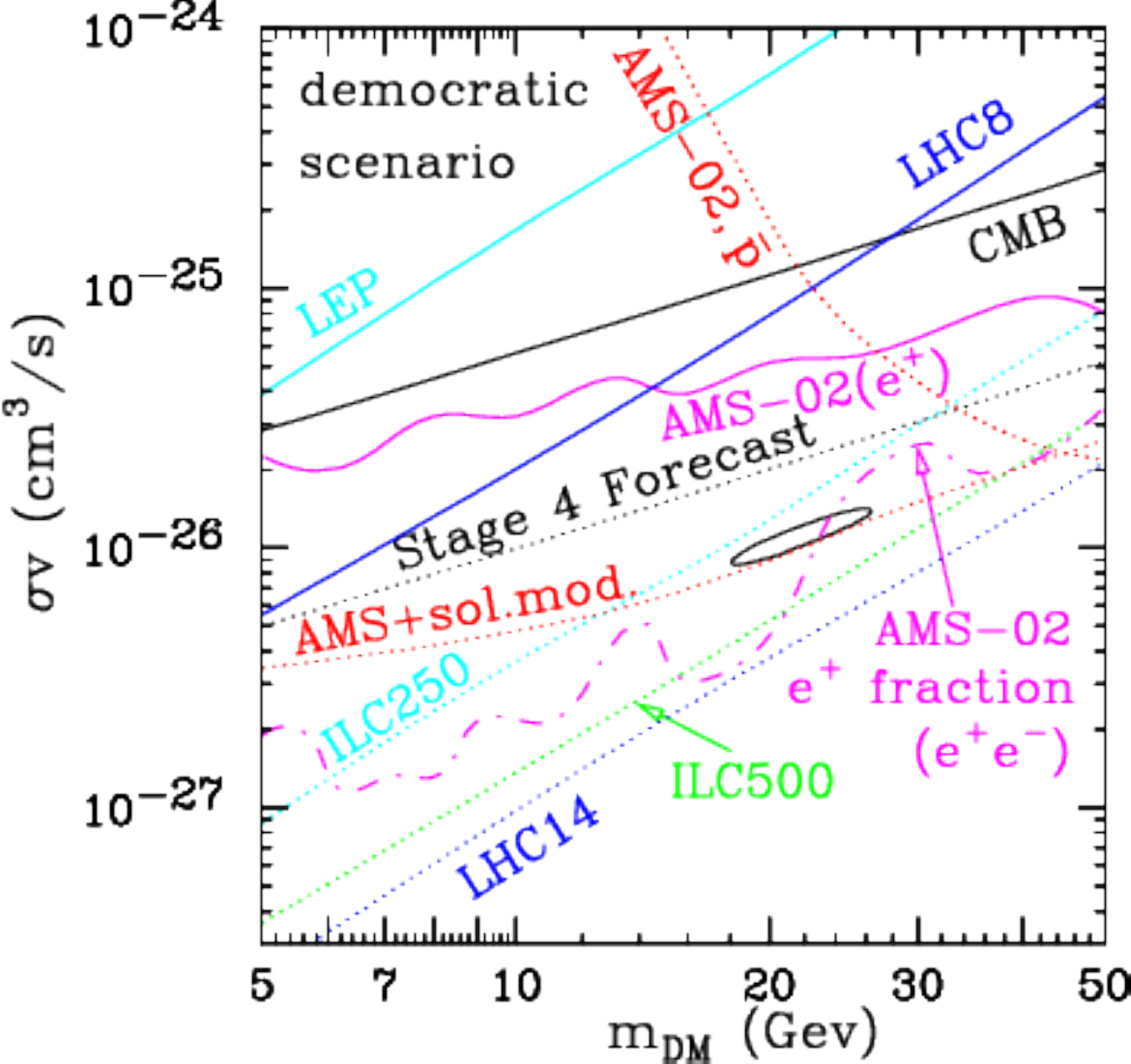}
\end{center}
\vspace*{-0.7cm}
\caption{Various constraints on dark matter annihilation cross section as a function of dark matter mass for the democratic scenario
as in Ref. \cite{HooperNew}.
}
\label{Fig2}
\end{figure}

Finally we would like to comment on other scenarios where dark matter annihilates to a combination of different channels.
First we consider a model in which the dark matter annihilates democratically to all kinematically accessible Standard Model fermions \cite{HooperNew}.
The fitted results are for 18-26 GeV for $m_{\rm DM}$ and (0.8-1.4) $\times 10^{-26} {\rm cm^3/s}$ for $\sigma v$,
which is between results in two previous cases, $b \bar b$ and $\ell\bar\ell$.
The corresponding 3$\sigma$ contour is shown in Figure \ref{Fig2}.
For the mass range of interest to fit the gamma-ray data, only charged leptons, neutrinos and five quark flavors contribute to the signal, and the corresponding annihilation fractions are $\frac{1}{7}$, $\frac{1}{7}$, and $\frac{5}{7}$, respectively.
Therefore for the electron final state, bounds from LEP/ILC and AMS get weaker by a factor of 21,
while those from LHC8/LHC14 (300 fb$^{-1}$) are rescaled by $\frac{5}{7}$.
However, for this democratic case, all five quark flavors contribute to DM production at the LHC, increasing sensitivity.
As a result, we find that LHC14 is the most powerful probe as shown in Figure \ref{Fig2}.
A future ILC with 500 GeV energy and a luminosity of 500 fb$^{-1}$ will be able to reach this model at a significance better than 3$\sigma$,
which is shown as a green line. A lower luminosity, 250 fb$^{-1}$, with a polarization of (+0.8, +0.5) would have 2-10 times better sensitivity
\cite{Chae:2012bq}.
Current and projected limits from CMB observations are obtained from the contribution for each channel which is weighted by the corresponding annihilation fraction and the efficiency factor $f_{\rm eff, sys}$.
Antiproton bounds are rescaled similarly.
As shown in Figure \ref{Fig2}, the democratic scenario is in severe tension with current AMS positron data, and future antiproton data and LHC14/ILC would rule out or confirm this scenario.

%%%%%%%%%%%%%%%%%%%%%%%%%%%%%%%%%%%%%%%%%%%%%%%%%%%%%%%%%%%%
%%%%%%%%%%%%%%%%%%%%%%%%%%%%%%%%%%%%%%%%%%%%%%%%%%%%%%%%%%%%

Another model is the case where the dark matter annihilates to a combination of channels, with cross sections proportional to the square of the mass of the final state particles.
In this case, the dominant channels are $b\bar b$, $c \bar c$, and $\tau^+\tau^-$.
With results in Figure \ref{Fig1}, no or mild, if any, bounds are anticipated from current experiments that we have discussed, although a dedicated study at LHC14 would be worth.

To conclude, Fermi-LAT gamma-ray excess may be explained by a relatively light dark matter in the mass range
where gauge boson final states are kinematically forbidden,
which implies that a hypothesized dark matter may annihilate into either leptonic or hadronic final states.
We find that AMS-02 positron and (future) antiproton data play a complementary role in constraining relevant parameter space or
excluding particular scenarios.
Similarly the LHC and a future ILC will look for different dark matter interaction in the mass range of interests.
The $\mu^+\mu^-$ ($b \bar b$) final state is also constrained by positron (antiproton) data as well.
The pure $\tau^+\tau^-$ final state is least constrained but there is mild tension with positron data and radio emission.
We find that current constraints indicate that
a naive scenario with a democratic branching fraction is severely constrained and
that dark matter couplings to second and/or (especially) third generation of fermions are preferred in the light of Fermi-LAT gamma-ray excess.
We anticipate that antiproton data of AMS-02 and LHC14 will provide an important guidance in seeking a microscopic model for the dark matter annihilation.

\vspace{0.5 cm}
\begin{acknowledgements}
K.K. is supported by the U.S. DOE under Grant No. DE-FG02-12ER41809 and by the University of Kansas General Research Fund allocation 2301566.
J.C.P. is supported by Basic Science Research Program through the National Research Foundation of Korea funded by the Ministry of Education (2011-0029758) and (NRF-2013R1A1A2061561).
\end{acknowledgements}

\end{document}